\documentclass[%
 reprint,
 amsmath,amssymb,
 aps,
]{revtex4-2}

\usepackage{graphicx}
\usepackage{dcolumn}
\usepackage{bm}

\begin{document}


\title{Analysis of $(3+1)D$ and $(2+1)D$ nonlinear ultrasonic waves using conformal invariance}

\author{Sadataka Furui}
 \email{furui@umb.teikyo-u.ac.jp}
\affiliation{Faculty of Science and Engineering, Teikyo University, 2-17-12 Toyosatodai, Utsunomiya, 320-0003 Japan}%
 
\author{ Serge Dos Santos}%
 \email{serge.dossantos@insa-cvl.fr}
\affiliation{ INSA Centre Val de Loire; INSERM, Imaging Brain \& Neuropsychiatry iBraiN U1253,  F-41034 Blois Cedex, France
}%

\date{\today}

\begin{abstract}
Localization and classification of scattered nonlinear ultrasonic signatures in 2 dimensional complex damaged media using Time Reversal based Nonlinear Elastic Wave Spectroscopy (TR-NEWS) approach is extended to 3 dimensional complex damaged media. In (2+1)D, i.e. space 2 dimensional time 1 dimensional spacetime, we used quaternion bases for analyses, while in (3+1)D, we use biquaternion bases.

The optimal weight function of the path of ultrasonic wave in (3+1)D lattice is obtained by using the Echo State Network (ESN) which is a Machine Learning technique. 
The hysteresis effect is incorporated by using the Preisach-Mayergoyz model.

We analyze the spectrum data of Wire Arc Additive Manufacturing (WAAM) sample obtained by Quaternion Excitation Symmetry Analysis Method (QESAM) using the conformally invariant quantum mechanical variables of de Alfaro-Fubini-Furlan and their supersymmetrically extended variables of Fubini-Rabinovici.  
\end{abstract}

\keywords{Nonlinear Elastic Wave Spectroscopy, Biquaternion, Quaternion, Conformal Symmetry}
\maketitle


\section{Introduction}
In (2+1)D image processing, Time Reversal based Nonlinear Elastic Wave Spectroscopy (TR-NEWS) was successful which is based on quaternions\cite{DSSF24,SFDS23a,DSCSS06,DSSF16}.  
Details of TR-NEWS experiment of detecting clacks in a block of layered Carbon Fibre Reinforced Polymer (CFRP) and 2D simulation using Newmark algorithm are given in \cite{LSDS17}. From differences of outputs from positive input and from negative inputs of receivers positioned at different positions, the position of a clack in the block was estimated.
In a paper of Fink's group\cite{MRDNF01}, cross-correlation of Time Reversal Mirror (TRM) technique\cite{BLEFF16} was presented. 

In (3+1)D image processing, the mapping by quaternion is replaced by that of biquaternion $M_2({\bf H})$ which is written in the text book of Garling\cite{Garling11}
\begin{equation}
j ({\mathcal A}^+_{3,1})=\left(\begin{array}{cc}
a_1 +a_2{\bf k}& b_1{\bf i}+b_2{\bf j}\\
c_1{\bf i}+c_2{\bf j}&d_1+d_2{\bf k}\end{array}\right),\nonumber
\end{equation}
where $a_i,b_i,c_i ,d_i\quad( i=1,2)$ are real and ${\bf i, j, k}$ are the pure quaternions.
The ultrasonic (US) wave is regarded as a phonon propagating in (3+1)D lattices filled with Weyl fermions represented by biquaternions.
We modify the fixed point (FP) action of DeGrand et al.\cite{DGHHN95} used in the simulation of Quantum Chromo Dynamics (QCD) by replacing Dirac fermions to Weyl fermions. The paths in (3+1)D consist of 1) A-type: paths on a 2D plane, 2) B-type: paths on 2 2D parallel planes expanded by ${\bf e}_1, {\bf e}_2$ connected by two links ${\bf e}_1\wedge {\bf e}_2$ and ${\bf e}_2\wedge{\bf e}_1$, 3) C-type: paths expanded by ${\bf e}_1,{\bf e}_2,{\bf e}_3$ and ${\bf e}_4$, which contain time delay or hysteresis effects.
Actions of A-type and B-type aree presented in \cite{SFDS23a,SFDS23b}, and we consider in this work, C-type paths.

The total length of a path is restricted to be less than or equal to 8 lattice units.  Since we ignore the path that runs along the periphery of a plaquette twice, there are 7 C-type paths. $L19,L20,L21,L22,L23,L24,L25$ in the notation of \cite{DGHHN95}.

Structure of this presentation is as follows. In section II, we show the paths in (3+1)D spacetime which is optimized by ESN. In section III, we present how to incorporate time delay or hysteresis effect in lattice simulations.
The Monte Carlo simulation result is shown in section IV. Hysteresis effects and Feynman's path integral are discussed in section V.
Quaternions appear in the framework of noncommutative geometry. Noncommutative geometry and quaternion Fourier transform are reviewed in section VI. Analysis of (2+1)D data using conformal symmetry is given in section VII.
Conclusion and outlook are presented in section VIII.

\section{The paths in (3+1)D spacetime}
In the Table 1, direction of paths in the biquaternion basis ${\bf e}_i {\bf e}_j$ and $x,y,z,t$ bases of 8 steps followed by TR 8 steps are shown. In the biquaternion basis, $i$ or $j$ of a step is chosen to be equal to that of the precedent and that of the subsequent step. The time step are ${\bf e}_1{\bf e}_4,{\bf e}_2 {\bf e}_4$ or ${\bf e}_3{\bf e}_4$.  When a step of $t$ or $-t$ are fixed to be ${\bf e}_k{\bf e}_4$, we fix the partner of $t$ or $-t$ to be the same. Except $L21,L22$, $k$ is uniquely fixed.
\begin{table*}[ht]
\begin{center}
\caption{Directions of the wave front of the C-type paths. The first line is in $R^4$ basis, the second line is in biquaternion basis.  }
{\small
\begin{tabular}{r|cccccccccccccccc}
 step & 1&2&3 &4&5&6&7&8&9&10&11&12&13&14&15&16\\
\hline
L19&x&y&z&t&-z&-t&-x&-y&-x&-y&-z&-t&z&t&x&y\\
&23&31&12&24&-12&-24&-23&-31&-23&-31&-12&-24&12&24&23&31\\
\hline
L20&x&y&z&t&-z&-y&-x&-t&-x&-y&-z&-t&z&y&x&t\\
&23&31&12&24&-12&-31&-23&-24&-23&-31&-12&-24&12&31&23&24\\
\hline
L25&x&y&z&t&-x&-y&-z&-t&-x&-y&-z&-t&x&y&z&t\\
&23&31&12&24&-23&-13&-12&-24&-23&-31&-12&-24&23&13&12&24\\
\hline
L21 &x&y&z&t&-z&-x&-t&-y&-x&-y&-z&-t&z&x&t&y\\
 &23&31&12&14/24&-12&-23&-34&-13&-23&-31&-12&-14/24&12&23&34&13\\
\hline
L22&x&y&z&t&-z&-x&-y&-t&-x&-y&-z&-t&z&x&y&t\\
&23&31&12&14/24&-12&-23&-31&-34&-23&-31&-12&-14/24&12&23&31&34\\
\hline
L23&x&y&z&t&-y&-x&-t&-z&-x&-y&-z&-t&y&x&t&z\\
&23&31&12&14&-31&-23&-24&-12&-23&-31&-12&-14&31&23&24&12\\
\hline
L24&x&y&z&t&-y&-x&-z&-t&-x&-y&-z&-t&y&x&z&t\\
&23&31&12&14&-31&-23&-12&-24&-23&-31&-12&-14&31&23&12&24\\
\end{tabular}
}
\end{center}
\end{table*}
In ESN there are recurrent layers of nonlinear units and linear, memory-less read-out layers which are trained. The matrices of input to output $W_{io}$ and input to reservoir $W_{ro}$ are fixed in \cite{Bianchi17}, however in our case, input to output $W_{io}$ is produced from $W_{ro}$.
\begin{figure*}[htb]
\begin{minipage}{0.47\linewidth}
\begin{center}
\includegraphics[width=6cm]{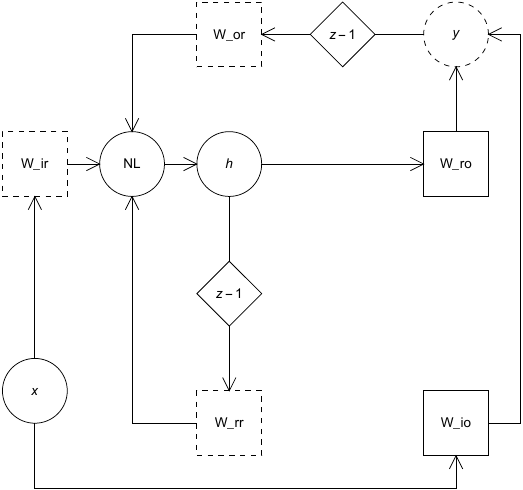}
\end{center}
\end{minipage}
\quad 
\begin{minipage}{0.47\linewidth}
\begin{center}
\includegraphics[width=6cm]{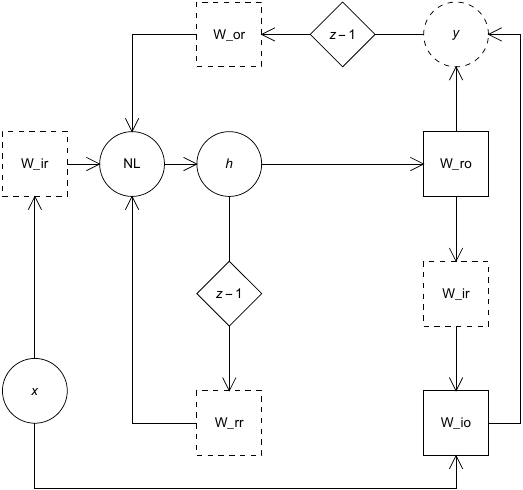}
\end{center}
\end{minipage}
\caption{Standard ESN (left) and our scheme of ESN. $W_{io}$ is produced by a multiplication of $W_{ir}$ on $W_{ro}$.(right) }.\label{ESN2}
\end{figure*}

\begin{figure*}[htb]
\begin{minipage}{0.47\linewidth}
\begin{center}
\includegraphics[width=3cm]{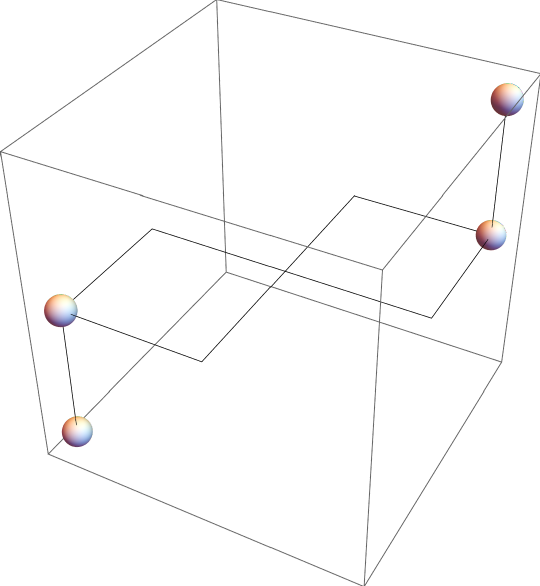}
\end{center}
\end{minipage}
\quad
\begin{minipage}{0.47\linewidth}
\begin{center}
\includegraphics[width=3cm]{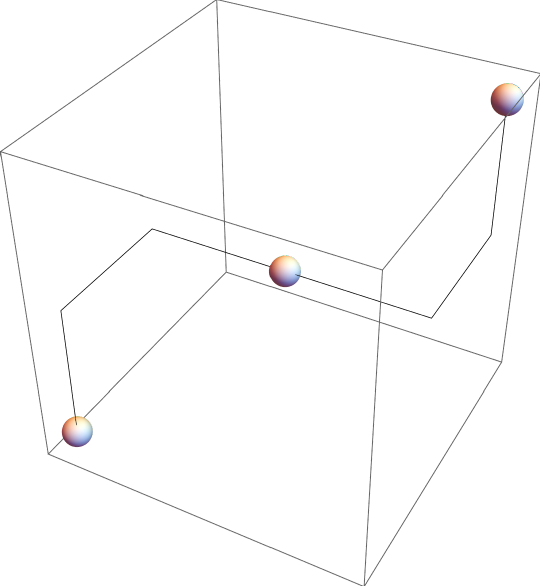}
\end{center}
\end{minipage}
\caption{The path of $L19$(left) and that of $L20$(right). Balls are the places where hysteretic time shift occurs. }\label{l1920}
\begin{minipage}{0.47\linewidth}
\begin{center}
\includegraphics[width=3cm]{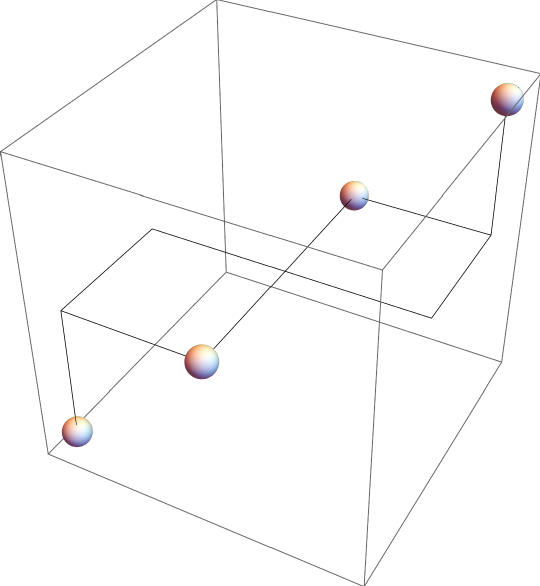}
\end{center}
\end{minipage}
\quad
\begin{minipage}{0.47\linewidth}
\begin{center}
\includegraphics[width=3cm]{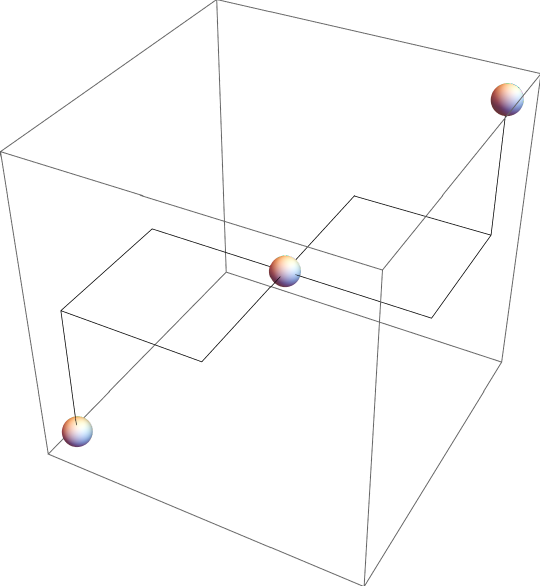}
\end{center}
\end{minipage}
\caption{The path of $L21$(left) and that of  $L22$(right).}\label{l2122}
\end{figure*}

\begin{figure*}[htb]
\begin{minipage}{0.47\linewidth}
\begin{center}
\includegraphics[width=3cm]{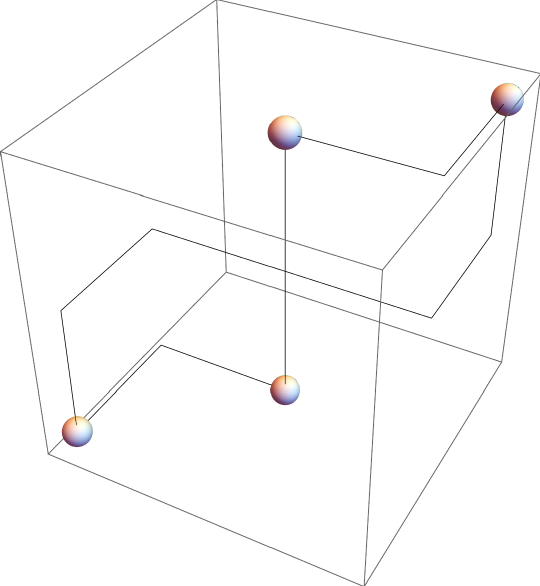}
\end{center}
\end{minipage}
\quad
\begin{minipage}{0.47\linewidth}
\begin{center}
\includegraphics[width=3cm]{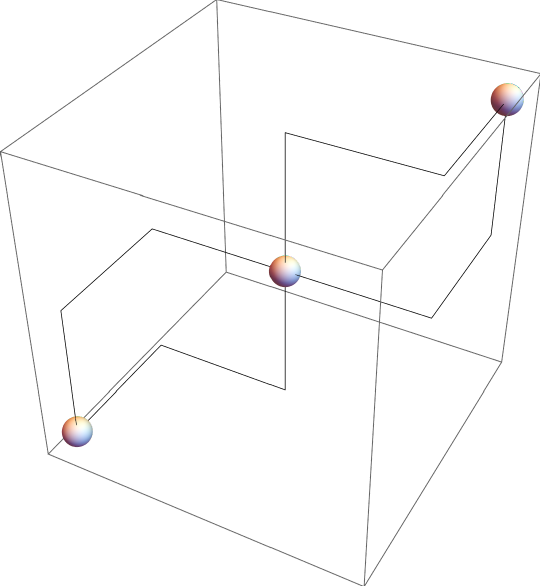}
\end{center}
\end{minipage}
\caption{The path of $L23$(left) and that of $L24$(right).}\label{l2324}
\begin{center}
\includegraphics[width=3cm]{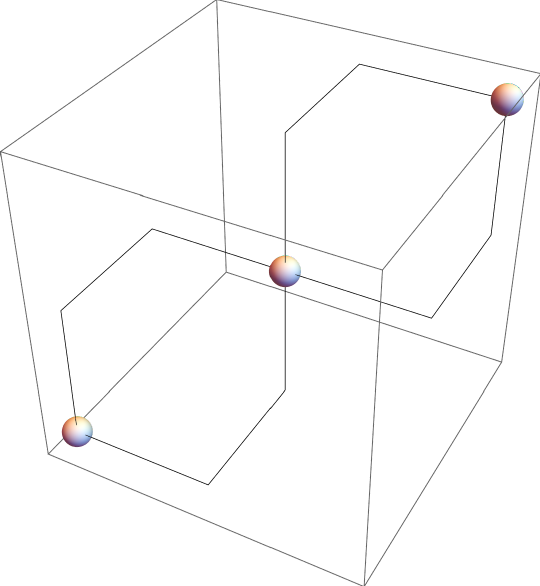}
\end{center}
\caption{The path of $L25$.}\label{l25}
\end{figure*}

The 16 steps of $L19,L20$, $L21,L22$, $L23,L24$ and $L25$ are shown in Fig.\ref{l1920} , Fig. \ref{l2122}, Fig.\ref{l2324} and Fig.\ref{l25}, respectively.  
At balls, time shifts occur. We assume same hysteretic effects occur stochastically in the balls.

The path of $L19, L21, L22$ differ by the points where time shifts  occur, We selected the time shift of $L19$ and $L20$ to be 24 and consider an average of 14 and 24 for $L21$ and $L22$. 

\section{Optimization of the weight function by ESN}
Since the path of $L19,L20,L21,L22,L23,L24,L25$ on the (3+1)D lattice are fixed\cite{DGHHN95}, ${\bf x}[t]$ for the seven paths can be fixed, when the 2D plane of the initial path is selected. Although ${\bf x}[t]$ runs in the 4D space, $W_{ir}{\bf x}[t]$ is a 7 dimensional vector defined on the path at time $t$.

We prepare random matrices
\begin{itemize}
\item  $W_{ro}$ $7\times 6$ matrix, Matrix elements obtained from the FP action..
\item  $W_{io}$ $4\times 6$ matrix. Matrix elements are fixed from $W_{ro}$ and $W_{ir}$.
\item  $W_{ir}$ $4\times 7$ matrix  Matrix elements are subject to be trained.
\item  $W_{or}$ $6\times 7$ matrix. Matrix elements are subject to be trained.
\item  $W_{rr}^{(t)}$ $7\times 7$ matrix. Matrix elements are subject to be trained. Mixing of paths at $t=4,7,8,12,15,16$ have specific filters.
\end{itemize}

By the criterion that at $t=4,7,8,12,15,16$, the paths which don't have the time shift are not disturbed, we multiply a filter to $W_{rr}$ such that non disturbed matrix elements are set to 0.  The interactions between reservoirs at these times are given in the following form.    
\begin{table*}[htb]
\caption{The ${\bf W}_{rr}^{(t)}$ matrices.}
\[
{\bf W}_{rr}^{(4)}=\left(\begin{array}{ccccccc}
 *&*&*&*&0&0&*\\
 *&*&*&*&0&0&*\\
 *&*&*&*&0&0&*\\
 *&*&*&*&0&0&*\\
 0&0&*&*&*&*&0\\
 0&0&*&*&*&*&*\\
 *&*&*&*&0&*&*\end{array}\right),\quad
 {\bf W}_{rr}^{(7)}=\left(\begin{array}{ccccccc}
 0&0&0&0&0&0&0\\
 0&0&0&0&0&0&0\\
 0&0&*&0&*&0&0\\
 0&0&0&0&0&0&0\\
 0&0&*&0&*&0&0\\
 0&0&0&0&0&0&0\\
 0&0&0&0&0&0&0\end{array}\right),
\quad
{\bf W}_{rr}^{(8)}=\left(\begin{array}{ccccccc}
 0&0&0&0&0&0&*\\
 0&*&0&*&0&0&*\\
 0&0&0&0&0&0&*\\
 0&*&0&*&0&0&*\\
 0&0&0&0&0&0&0\\
 0&0&0&0&0&*&*\\
 *&*&*&*&0&*&*\end{array}\right),
\]
\quad
\[
{\bf W}_{rr}^{(12)}=\left(\begin{array}{ccccccc}
 *&*&*&*&0&0&*\\
 *&*&*&*&0&0&*\\
 *&*&*&*&*&*&*\\
 *&*&*&*&*&*&*\\
 0&0&*&*&*&*&0\\
 0&0&*&*&*&*&*\\
 *&*&*&*&0&*&*\end{array}\right),
\quad 
{\bf W}_{rr}^{(15)}=\left(\begin{array}{ccccccc}
 0&0&0&0&0&0&0\\
 0&0&0&0&0&0&0\\
 0&0&*&0&*&0&0\\
 0&0&0&0&0&0&0\\
 0&0&*&0&*&0&0\\
 0&0&0&0&0&0&0\\
 0&0&0&0&0&0&0\end{array}\right),\quad
{\bf W}_{rr}^{(16)}=\left(\begin{array}{ccccccc}
 0&0&0&0&0&0&0\\
 0&*&0&*&0&*&*\\
 0&0&0&0&0&0&0\\
 0&*&0&*&0&0&0\\
 0&0&0&0&0&0&0\\
 0&*&0&0&0&*&*\\
 0&*&0&0&0&*&*\end{array}\right).
\]
\end{table*}

The low and the column are in the order $L19,L20,L21,L22,L23,L24,L25$ and $*$ in matrices of $W_{rr}^{(t)}$ indicates time shift in the own path occur or time shifts that cause mixing of paths occur. 

We prepare 40 random  vectors for $t=4 :{\bf h}[4]=f(W_{ir}{\bf x}[4])=( {h}_1[4], {h}_2[4],\cdots, {h}_7[4])$ corresponding to $L19,L20,\cdots , L25$. Here $f$ is the nonlinear activation function, ${\bf x}[4]=({x}_1[4],{x}_2[4],{x}_3[4],{x}_4[4])$, and $W_{ir}({x}_1[4],{x}_2[4],{x}_3[4],{x}_4[4])$ is the seven dimentional vector given by the fixed point action of DeGrand et al.\cite{DGHHN95}.  The matrix $W_{io}$ is normalized such that $\sum_{i=1}^4 \sum_{j=1}^6 (W_{io})_{i,j}^2=\sum _{r=1}^7\sum_{j=1}^6 (W_{ro})_{r,j}^2$.

We define the loss function ${\mathcal L}=S_{ro}^2$, where $S_{ro}=W_{ro}{\bf h}-{\bf y}^*$. In the calculation of ${\bf h}[t]$, we choose
\begin{equation}
{\bf h}[t]=f(W_{io}{\bf x}[t]+W_{ro}({\bf h}[t]+{\bf b}_h)),
\end{equation}
where 
\begin{equation}
{\bf b}_h=-2\eta\frac{\partial {\mathcal L}}{\partial S_{ro}}\frac{\partial S_{ro}}{\partial h}=-2\eta(W_{ro}{\bf h}-{\bf y}^*)W_{ro}.
\end{equation}

The weight matrix of reservoirs ${W}_{rr}^{(t)}$ can also be trained to yield proper outputs. We assume that the matrix element of ${W}_{rr}^{(t)}$, denoted as $W_{m,n}^{(t)}$ is not 0, when the direction of the path $m$: $e_i e_4$ and that of $n$ : $e_j e_4$ coincide. At $t=4,7,8,12,15$ and 16, there are time shift points between different paths.   Time shifts occur with biquaternion base 14, 24 or 34, and when the shift of different paths have the same biquaternion base, we assume that mixings of paths occur and the matrix element differs from 0. 

Optimiztion steps for $W_{rr}^{(t)}$ are done by choosing as an activation function the sigmoid function
\begin{equation}
 \frac{1}{1+exp(-W_{rr}^{(t)}{\bf h}[t]-W_{ir}{\bf x}[t+1]-W_{or}{\bf y}[t])}
\end{equation}
which has the range $(0,1]$ and its derivative with respect to $W_{rr}^{(t)}$ as
\begin{eqnarray}
&&\frac{1-\frac{1}{1+exp(-W_{rr}^{(t)}{\bf h}[t]-W_{ir}{\bf x}[t+1]-W_{or}{\bf y}[t])}}{1+exp(-W_{rr}^{(t)}{\bf h}[t]-W_{ir}{\bf x}[t+1]-W_{or}{\bf y}[t])}
{\bf h}[t]\nonumber\\
&&=D W_{rr}^{(t)}{\bf h}[t],
\end{eqnarray}
or the $\tanh$ functon $\tanh(-W_{rr}^{(t)}{\bf h}[t]-W_{ir}{\bf x}[t+1]-W_{or}{\bf y}[t])$ which has the range $[-1,1]$ and its derivative with respect to $W_{rr}^{(t)}$ is
\begin{equation} 
-{\rm sech}^2(-W_{rr}^{(t)}{\bf h}[t]-W_{ir}{\bf x}[t+1]-W_{or}{\bf y}[t]){\bf h}[t]=DW_{rr}^{(t)}{\bf h}[t].,
\end{equation}
$W_{rr}^{(t)}$ is modified to
\begin{equation}
W_{rr}^{(t)}-\eta\frac{\partial{\mathcal L}}{\partial W_{rr}^{(t)}}=W_{rr}^{(t)}-\eta D W_{rr}^{(t)}.
\end{equation}
The optimization of $W_{rr}^{(t)}$ and the state vector ${\bf h}[t]$ are done as in Appendix.

The calculation of ${\bf h}[4]$ changes from the first cycle to ${\bf h}[4']=f(W_{rr}{\bf h}[3]+W_{i r}{\bf x}[4]+W_{or}{\bf y}[3])$.
The cycle from 4 to 20 continues until optimized $W_{ir}, W_{rr}, W_{or}$ are found.

Store ${\bf S}=\left[\begin{array}{cc}
{\bf x}^T[4]& {\bf h}^T[4]\\
{\bf x}^T[7]& {\bf h}^T[7]\\
{\bf x}^T[8]& {\bf h}^T[8]\\
{\bf x}^T[12]& {\bf h}^T[12]\\
{\bf x}^T[15]& {\bf h}^T[15]\\
{\bf x}^T[16]&{\bf h}^T[16]
\end{array}\right]$ and ${\bf y}^*=\left[\begin{array}{c}
{\bf y}^*[4]\\
{\bf y}^*[7]\\
{\bf y}^*[8]\\
{\bf y}^*[12]\\
{\bf y}^*[15]\\
{\bf y}^*[16]
\end{array}\right]$.

As in the case of $(2+1)D$, we optimize the weight function of 7 paths ($L19,L20,L21,L22,L23, L24, L25$) in $(3+1)D$ that minimize the loss ${\mathcal L}=||{\bf S}\,{\bf W}-{\bf y}^*||^2$, where ${\bf W}=[W_{io}, W_{ro}]^T$, and.
\begin{equation}
{\bf S}{\bf W}=\left[\begin{array}{cc}
(W_{io}{\bf x})^T[4]&(W_{ro}{\bf h})^T[4]\\
(W_{io}{\bf x})^T[7]&(W_{ro}{\bf h})^T[7]\\
(W_{io}{\bf x})^T[8]&(W_{ro}{\bf h})^T[8]\\
(W_{io}{\bf x})^T[12]&(W_{ro}{\bf h})^T[12]\\
(W_{io}{\bf x})^T[15]&(W_{ro}{\bf h})^T[15]\\
(W_{io}{\bf x})^T[16]&(W_{ro}{\bf h})^T[16]
\end{array}\right].
\end{equation}
We adopt a cylindrical lattice model, such that 7 paths start from the origin of a space and returns to the origin. The total action becomes 0 when the path returns to the origin. Therefore at $t=8$ and $t=16$,  $W_{ir}{\bf x}[t]$ become 0 at these epochs.

We are considering paths in momentum space, the Fourier transform of the path in the position space. In ${\bf x}[t]$ of the calculation we consider the projection space $RP^4$ which means that we optimize the scale of the 4D vectors.

We can distinguish the path of time shift through ${e}_1 {e}_4$ and through ${e}_2 {e}_4$, and consider 9 dimensional bases and perform the Elman Recurrent Neural Network (ERNN)\cite{Bianchi17,BSURS15}. 

\section{Monte Carlo simulation results}
The searched output ${\bf y}[t]$ of ESN, which is 7 dimensional vector, and $t$ runs from $t=4,\cdots 16, 1, 2, 3$ modulus 16. We want to obtain weights of 7 bases which are stable over many cycles. We observed that for the nonlinear activation function, $\tanh$ is better than the sigmoid function. 

The $\tanh$ function was used in the Long Short Term Memory (LSTM) cells method\cite{RLM22}.

The deviations $||{\bf y}[t]-{\bf y}^*[t]||^2$ using the $\tanh$ function are much reduced from those obtained by the sigmoid function. However, near $t=9$ and $t=16$, large deviation appear. It is due to the fact that at $t=8$ and $t=16$, the random walk returns to the original position, and  $y^*[t]$ is taken to be zero. The good property of this method is that from the 3000th cycle to the 4000th cycle, the weight is stable. After 5000 cycles, the absolute value of the correlation increases slghtly. 

The obtained ${\bf y}[9]$ and ${\bf y}[16]$ suggests that ${\bf y}^*[8]={\bf y}^*[16]=0$ derived from the action of paths returning to the original in 3D space meight be inappropriate. 

 In order to visualize the stability of outputs, we calcurated the correlation of $H=W_{ro}{\bf h}$ and $Y={\bf y}$ both $7\times 7$ matrices
 which is expressed as
 \begin{equation}
 \left(\begin{array}{cccccc}
\frac{\sigma_{H_1Y_1}}{\sigma_{H_1}\sigma_{Y_1}}&\frac{\sigma_{H_1Y_2}}{\sigma_{H_1}\sigma_{Y_2}}&\cdots&\frac{\sigma_{H_1Y_5}}{\sigma_{H_1}\sigma_{Y_5}}&\frac{\sigma_{H_1Y_6}}{\sigma_{H_1}\sigma_{Y_6}}& \frac{\sigma_{H_1 Y_7}}{\sigma_{H_1}\sigma_{Y_7}}\\
\vdots&\vdots&\vdots&\vdots&\vdots\\
\frac{\sigma_{H_7 Y_1}}{\sigma_{H_7}\sigma_{Y_1}}&\frac{\sigma_{H_7 Y_2}}{\sigma_{H_7}\sigma_{Y_2}}&\cdots&\frac{\sigma_{H_7 Y_5}}{\sigma_{H_7}\sigma_{Y_5}}&\frac{\sigma_{H_7 Y_6}}{\sigma_{H_7}\sigma_{Y_6}}&\frac{\sigma_{H_7,y_7}}{\sigma_{H_7}\sigma_{Y_7}}\end{array}\right)
 \end{equation}

\begin{figure*}[htb]
\begin{minipage}{0.47\linewidth}
\begin{center}
\includegraphics[width=6cm]{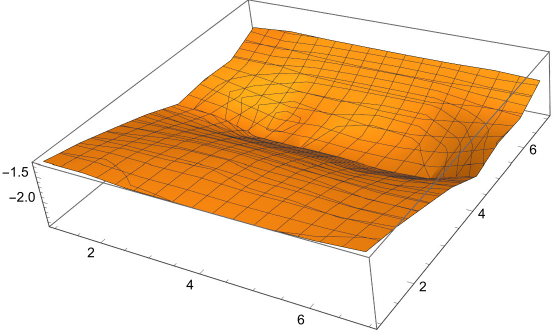}
\end{center}
\end{minipage}
\quad
\begin{minipage}{0.47\linewidth}
\begin{center}
\includegraphics[width=6cm]{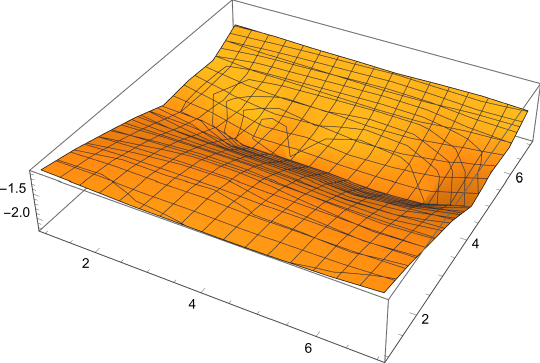}
\end{center}
\end{minipage}
\caption{The correlation function of $W_{ro}{\bf h}$ and $\bf y$ at the 3000th cycle (left), at the 4000th cycle (right),  }\label{gCor}
\end{figure*}

 From the 3000th cycle to 4000th cycle,  $\frac{\sigma_{H_i Y_j}}{\sigma_{H_i}\sigma_{Y_j}}$ multiplied by 100 are plotted in Figure\ref{gCor}.
Correlations at the 3000th cycle and at the 4000th cycle are similar. As the number of cycles increases, the depth of correlation surface becomes deeper.

We tried to run the program up to 10000 cycles, but the Mathematica produced a warning that the accuracy may be decreased. Therefore, we consider data between 3000 cycles and 4000 cycles and abort our choice of ${\bf y}^*[t]$, and  took ${\bf y}^*[t]=0$ for all $t$.

\begin{figure*}[htb]
\begin{minipage}{0.47\linewidth}
\begin{center}
\includegraphics[width=8cm]{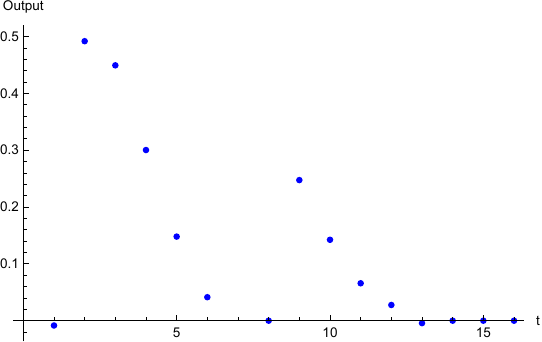}
\end{center}
\end{minipage}
\quad
\begin{minipage}{0.47\linewidth}
\begin{center}
\includegraphics[width=8cm]{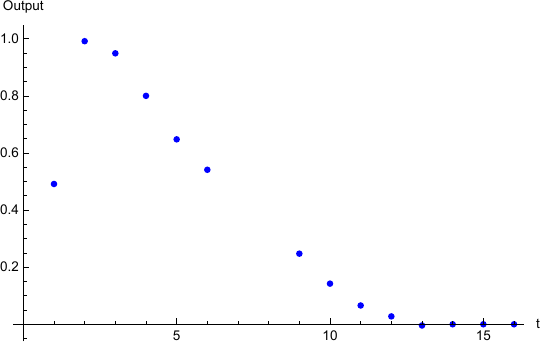}
\end{center}
\end{minipage}
\caption{Output calculated by the weight function at the 4000th cycle and the Fixed Point action at $t=2,\cdots, 16$ (left).
The output in the range $0<t<8$ shifted upward by 0.5. (right)}\label{output}
\end{figure*}

The obtained weight function is multiplied by the fixed point action at $t=T/16, T=0,\cdots, 255$ which is used in the previous calculation\cite{SFDS23b}. Actions of $L21$ and $L22$ are chosen to be the average of $e_1e_4$ and $e_2e_4$ contributions. The actions at $t=8$ and $t=16$ are 0, and the output in the range $9\leq t\leq 12$ and $4\leq t\leq 6$ are almost parallel.

We shifted the output in the range $0<t<8$ of all cycles by adding 0.5, and found that outputs in $0<t<8$ and $8<t<16$ become smooth. 
The shift of action in $0<t<8$ can be interpleted as an additional action of Preisach-Mayergoyz (PM) model\cite{Mayergoyz03} which is discussed in the next section.

\section{Hysteresis effects and Feynman's path integral}
Feynman showed the probability amplitude for a space-time path in the region $R$ as $|\varphi(R)|^2$\cite{Feynman48}, where
\begin{equation}
\varphi(R)=\lim_{\epsilon\to 0}\int_R exp[\frac{\sqrt{-1}}{\hbar}\sum_i S(x_{i+1},x_i)]\cdots\frac{dx_{i+1}}{A}\frac{dx_i}{A}\cdots.
\end{equation}
Here $\frac{1}{A}$ is a normalization factor to make the wave function satisfies the Schr\"odinger equation. The probability that the path is in $R'$ and later in $R''$ is $|\varphi(R',R'')|^2$, where 
\begin{eqnarray}
\varphi(R',R'')&=&\int \chi^*(x,t)\psi(x,t) dx,\\
\psi(x_k,t)&=&\lim_{\epsilon\to 0}\int_{R'}exp[\frac{\sqrt{-1}}{\hbar}\sum_{t=-\infty}^{k-1}S(x_{i+1},x_i)]\nonumber\\
&&\times\frac{dx_{k-1}}{A}\frac{d x_{k-2}}{A}\cdots,\\
\chi^*(x_k,t)&=&\lim_{\epsilon\to 0} \int_{R''} exp[\frac{\sqrt{-1}}{\hbar}\sum_{i=k}^\infty S(x_{i+1}, x_i)]\nonumber\\
&&\times\frac{dx_{k+1}}{A}\frac{dx_{k+2}}{A}\cdots.
\end{eqnarray}

The wave function at $t=t+\epsilon$ is approximated as
\begin{equation}
\psi(x_{k+1},t+\epsilon)=\int exp[\frac{\sqrt{-1}}{\hbar}S(x_{k+1}, x_k)]\psi(x_k,t) \frac{dx_k}{A}.
\end{equation}

The PM model\cite{Mayergoyz03} contains many building blocks of $\varphi(R', R'')$ with time delay. It is formulated by the input $u(t)$ and output $f(t)$ as\cite{PKDS12}
\begin{equation}
f(t)=\hat \gamma u(t)=\int\int_{\alpha\geq \beta}\mu(\alpha,\beta)\hat\gamma_{\alpha\beta} u(t) d\alpha d\beta,
\end{equation}
where 
\begin{equation}
\hat\gamma_{\alpha\beta} u(t)=\left\{ \begin{array}{cc}
1,& u(t)\geq \alpha,\\
0,& u(t)\leq \beta,\\
k,& u(t)\in(\beta,\alpha),\end{array}\right.
\end{equation}
\begin{equation}
k=\left\{\begin{array}{cc}
1&\exists t^*:u(t^*)>\alpha \quad {\rm and}\quad \forall\tau\in (t^*,t), u(\tau)\in (\beta,\alpha),\\
0&\exists t^*:u(t^*)<\beta \quad {\rm and}\quad \forall\tau\in (t^*,t),u(\tau)\in (\beta,\alpha).\end{array}\right.
\end{equation}

The PM model is constructed as a superposition of $\gamma_{\alpha\beta}u(t)$ multipied by the weight function $\mu(\alpha,\beta)$.

In magnetism, input $u(t)$ corresponds to the magnetic field $H(t)$, and $\gamma_{\alpha\beta}u(t)$ corresponds to the magnetization $M(t)$.

Hysteresis effects calculated by using quaternion basis differ from those of standard PM model. 
We also remark that the 4D echo technique in medical examinations is a record of 3D images in the time series, and hysteresis effects are not considered.

\section{Noncommutative geometry and quaternion Fourier transform}
 Quaternion and biquaternion basis model can be used not only for NDT, but also for QCD lattice simulations. Quaternion quantam Mechanics was proposed by Finkelstein et al.\cite{FJSS62} in 1962, and Adler\cite{Adler85,Adler94} used quaternions in generalized quantum dynmics, and in operator gauge invariant quaternionic field theory. He considered the total trace Lagrangean and Hamiltonian dynamics and asked, "Given two scalar or fermion quaternionic operator fields, is there a criterion for determining whether they are related by a bi-unitary operator gauge transformation?"\cite{Adler94}.
Our systems are not related by a gauge transformation but related by a different choice of bases.

Quantum mechanics represented by quaternions is proposed also by Connes\cite{Connes94}. 

In Heisenberg picture, the equation of motion is
\begin{equation}
\sqrt{-1}\hbar \partial_t\psi({\bf x},t)=[\psi({\bf x},t),H].
\end{equation}
When there are hysteresis effect, Connes extended the equation of motion using groupoids in dynamical systems expressed by $(X,R,\pi)$,
where $X$ is a topological space, $R$ is the real number space, and $\pi$ is a mapping from $X\times R$ to $X$. 

Algebraic structures of a Groupoid $G$ and its distinguished subset $G^{(0)}$ are characterised by the source map $s$ and the range map $r$ on an element $\gamma\in G$
\begin{equation}
(a*b)(\gamma)=\sum_{\gamma_1\circ \gamma_2=\gamma} a(\gamma_1)b(\gamma_2)
\end{equation}
where $a,b$ are arbitrary maps and
\begin{equation}
\circ: G^{(2)}=\{(\gamma_1,\gamma_2)\in G\times G; s(\gamma_1)=r(\gamma_2)\}\to G.
\end{equation}
The source map and the range map satisfy
\begin{enumerate}
\item $s(\gamma_1\circ\gamma_2)=s(\gamma_2), r(\gamma_1\circ \gamma_2)=r(\gamma_1)$.
\item $s(x)=r(x)=x, x\in G\to G^{(0)}$.
\item $\gamma\circ s(\gamma)=\gamma$,  $r(\gamma)\circ \gamma=\gamma$.
\item $(\gamma_1\circ\gamma_2)\circ \gamma_3=\gamma_1\circ(\gamma_2\circ\gamma_3)$.
\item $\gamma \gamma^{-1}=r(\gamma)$, $\gamma^{-1}\gamma=s(\gamma)$.
\end{enumerate}
We consider $G^{(0)}\subset M\times [0,1]$ with inclusion
\begin{eqnarray}
&&({\bf x},\epsilon)\to ({\bf x,x},\epsilon)\in M\times M\times [0,1]\quad {\rm for}\quad x\in M,\epsilon>0 \nonumber\\
&&({\bf x},0)\to {\bf x}\in M\subset TM,
\end{eqnarray}
where $TM$ is the tangent manifold defined by the sequence $(x_n,y_n,\epsilon_n)$ in $G_1=M\times M\times ]0,1]$ in the limit of
\begin{equation}
{\bf x}_n\to {\bf x},\quad {\bf y}_n\to {\bf y},\quad \frac{{\bf x}_n-{\bf y}_n}{\epsilon_n}\to X.
\end{equation}

The range map and the source map satisfy
\begin{equation}
\left\{\begin{array}{l}
r({\bf x,y},\epsilon)=(x,\epsilon) \quad {\rm for }\quad x\in M,\epsilon>0\nonumber\\
r({\bf x},X)=({\bf x},0)\quad {\rm for}\quad {\bf x}\in M, X\in T_x(M) \end{array}\right.
\end{equation}
\begin{equation}  
\left\{ \begin{array}{l}
s({\bf x,y},\epsilon)=(y,\epsilon) \quad {\rm for }\quad y\in M\epsilon>0\nonumber\\
s({\bf x},X)=({\bf x},0)\quad {\rm for}\quad {\bf x}\in M, X\in T_x(M). \end{array}\right.
\end{equation}
The composition is
\begin{eqnarray}
&&({\bf x,y},\epsilon)\circ({\bf y,z},\epsilon)=({\bf x,z},\epsilon) \quad{\rm for}\quad \epsilon>0\nonumber\\
&& \quad{\rm and}\quad {\bf x,y,z}\in M,\nonumber\\
&&({\bf x},X)\circ ({\bf x},Y)=({\bf x},X+Y)\quad {\rm for}\quad {\bf x}\in M\quad\nonumber\\
&&\quad{\rm and}\quad X,Y\in T_x(M).
\end{eqnarray}
\begin{figure}[htb]
\begin{center}
\includegraphics[width=6cm]{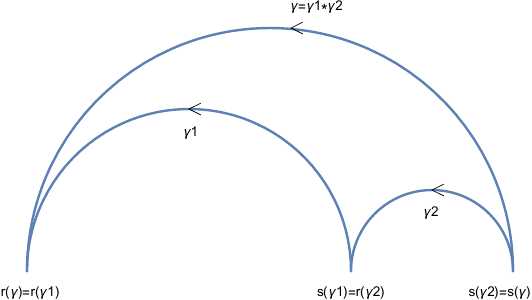}
\end{center}
\caption{The groupoid structure of the hysteresis effect. $r(\gamma)=({\bf x},0)$.}\label{Groupoid}
\end{figure}
To allow hysteresis effects we allow in the case of $\gamma=\gamma_1\circ\gamma_2$, $r(\gamma_2)=({\bf x},1)=s(\gamma_1)$, where 1 is in the unit of time shift in our FP lattice model, and $\bf x$ is the position of the ball in the Figure 3,4 and 5.
The PM model says that the output becomes a sum of Groupoids including time shifts
In the book of \cite{BP93} formulae of measurling cross-correlations in various setups are given. 
For input signal $x(t)$ and output signal $y(t)$ one uses the following quantities
\begin{itemize}
\item $x(t)=$ on going natural input (unmeasured).
\item $i(t)=$ known external input signal (measured)
\item $v(t)=$ linear output (unmeasured) caused by $x(t)$.
\item $r(t)=$ linear output (unmeasured) caused by $i(t)$.
\item $n(t)=$ unknown output noise (unmeasured).
\item $y(t)=v(t)+r(t)+n(t)=$ total output signal (measured).
\end{itemize}
Their Fourier transforms are related as
\begin{eqnarray}
&&R(f,T)=H(f)I(f,T), \quad V(f,T)=H(f)X(f,T)\nonumber\\
&&Y(f,T)=R(f,T)+V(f,T)+N(f,T)\nonumber\\
&&\quad \quad \quad=H(f)[I(f,T)+X(f,T)]+N(f,T)
\end{eqnarray}
where $H(f)$ is the frequency response function, and $T$ is the maximal time. 

The cross-spectrum terms $G_{ix}(f)=G_{in}(f)=G_{nv}(f)=G_{nr}(f)$ are assumed to be 0.
\begin{eqnarray}
&&G_{yy}(f)=G_{rr}(f)+G_{vv}(f)+G_{nn}(f)\nonumber\\
&&G_{iy}(f)=H(f)G_{ii}(f)
\end{eqnarray}
When the external excitation signal is a white noise $G_{ii}(f)=K$ and $H(f)=\frac{G_{iy}(f)}{K}$.
\section{Analysis of (2+1)D data using conformal symmetry}
The Time Reversal based Nonlinear Elastic Wave Spectroscopy (TR-NEWS) experiment on the WAAM sample is done by a ultrasonic wave transducer $T_x$ and a  receiver made of WAAM sample, which consists of 12 layers $R1_x,R2_x,\cdots ,R12_x$ placed on an edge of rectangular materials. See Fig.9. US waves produced by $T_x$ is scattered in the material and received by $R1_x,R2_x,\cdots,R12_x$.  The ultrasonic wave propagate on a 2 dimensional (2D) plane and taking into account the hysteresis effect, the experiment is done in (2+1)D spacetime.    
\begin{figure}[htb]
\begin{center}
\includegraphics[width=7cm,angle=0,clip]{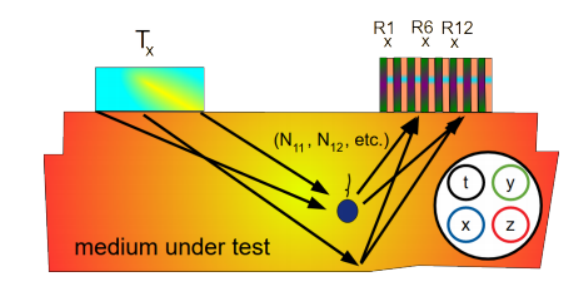}
\caption{The experimental setup of TR-NEWS using WAAM sample..}\label{TRNEWS12}
\end{center}
\end{figure}
In 1976, de Alfaro-Fubini-Furlan \cite{DAFF76} showed that the properties of a field theory in one overall time dimension and space 2 dimension can be studied by the $O(2,1)$ group with three generators : $H,D$ and $K$, where $H$ is Hamiltonian, $D$ is dilation genertor and $K$ is conformal transformation generator.
They considered the Lagrangian 
\begin{equation}
{\mathcal L}=\frac{1}{2}(\dot Q^2-\frac{g}{Q^2})
\end{equation}
and three generators $H,D,K$, which obey the algebra
\begin{equation}
[H,D]=\sqrt{-1} H,\quad [K,D]=-\sqrt{-1}K,\quad [H,K]=2\sqrt{-1} D.
\end{equation}
Any combination 
\begin{equation}
G=u H+v D+w K
\end{equation}
satisfy
\begin{equation}
\frac{\partial G}{\partial t}+\sqrt{-1}[H,G]=0.
\end{equation}
The algebra of $O(2,1)$ appears for
\begin{equation}
R=\frac{1}{2}(\frac{1}{a}K+a H),\quad S=\frac{1}{2}(\frac{1}{a}K-a H),
\end{equation}
as
\begin{equation}
[D,R]=\sqrt{-1}S,\quad [S,R]=-\sqrt{-1}D,\quad [S,D]=-\sqrt{-1}R.
\end{equation}
The operator $G$ is compact when $\Delta=v^2-4 u e>0$.

Fubini and Rabinovici \cite{FR84} extended the model of DFF by considering supersymmetric version of conformal QM.
The algebra is given by
\begin{equation}
\frac{1}{2}\{ Q,Q^\dagger\}=H,\quad \{ Q,Q \}=\{ Q^\dagger,Q^\dagger\}=0,
\end{equation}
where
\begin{equation}
Q=\psi^\dagger(-\sqrt{-1}p+\frac{dW}{dx}),\quad Q^\dagger=\psi(\sqrt{-1}p+\frac{dW}{dx}),
\end{equation}
$p=-\sqrt{-1}(\partial/\partial x)$ and $W(x)$ is a potential.  $\psi$ and $\psi^\dagger$ satisfy $\{ \psi ,\psi^\dagger \}=1$ and
$\frac{1}{2}[\psi^\dagger,\psi]=B$. The latter is the generator of U(1) transformation $\psi\to e^{\sqrt{-1}\alpha}\psi$, $\psi^\dagger\to e^{-\sqrt{-1}\alpha}\psi^\dagger$, with eigenvalues 1/2 and -1/2 respectively.

The Hamiltonian $H$ of DFF satisfy
\begin{equation}
\frac{1}{2}(H K+K H)-D^2=\frac{g}{4}-\frac{3}{16}.
\end{equation}
Using $H K-K H=2\sqrt{-1}D$, we obtain $H K-D^2+\sqrt{-1}D=\frac{g}{4}-\frac{3}{16}$

We take an eigenstate $\psi$ and $K\psi=\chi$
\begin{eqnarray}
\langle \chi | H K|\psi\rangle&=&\langle \chi|H|\chi\rangle=\langle \chi| D^2-\sqrt{-1}D+\frac{g}{4}-\frac{3}{16}| \psi\rangle\nonumber\\
&=&\langle\chi |D^2-\sqrt{-1}D|\psi\rangle+\langle\chi|\frac{g}{4}-\frac{3}{16}|\psi\rangle
\end{eqnarray}
When one takes the real expectation value and $D^2$ is the Casimir operator, we may be allowed to approximate the right hand side as $\langle\chi|D^2|\psi\rangle\sim \langle \chi|D^2|\chi\rangle=J(J-1)$

\begin{figure}[htb]
\begin{center}
\includegraphics[width=7cm]{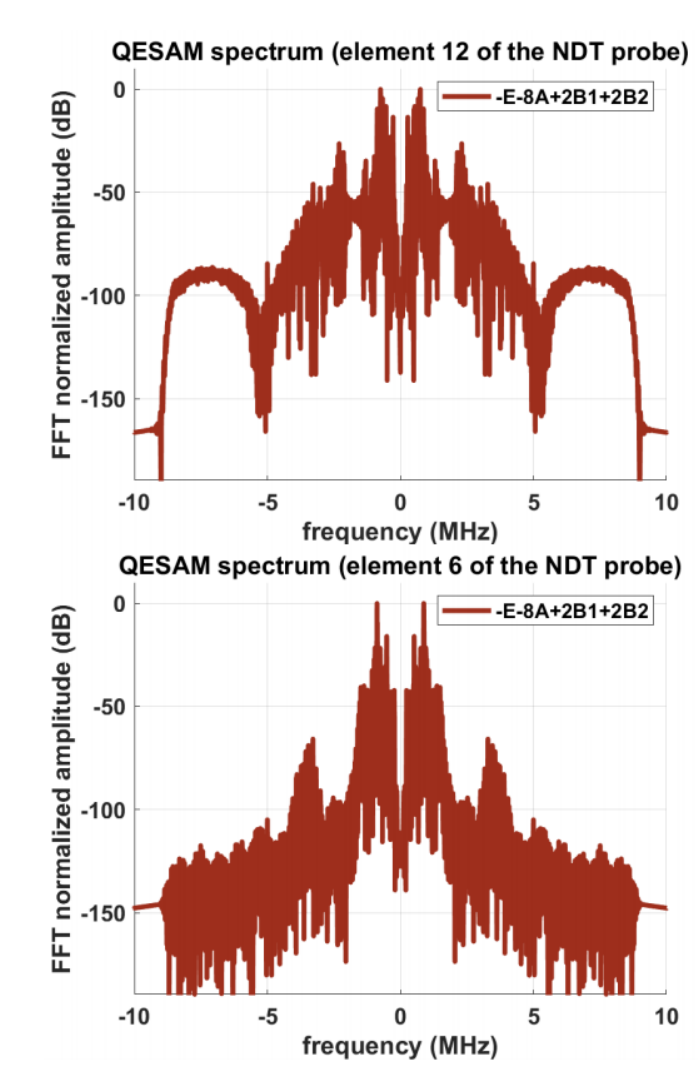}
\caption{The spectra of QESAM received at R12(top) and at R6(bottom).}\label{QESAM}
\end{center}
\end{figure}

The peaks of QESAM spectrum or $R12_x$ appear at $2\Delta_\nu, 6\Delta_\nu$ and the dip of the spectrum appears at $13\Delta_\nu, 23\Delta_\nu$ 
($\Delta_\nu$ corresponds to 0.55MHz in the Fig \ref{QESAM}.)

For $R6_x$, peaks appear at $3\Delta_\nu, 9\Delta_\nu$ and the dips are outside the range of measurement.

The coefficient of peaks $2=2*1$ and $6=3*2$ are expection value of the Casimir operator.

The coefficient of dips 13 is close to $4*3+1$ but 23 may be an artifact near the edge of spectrum.
In conformal supersymmetric model with a choice of parameter $a=1$, the eigenvalue of $G=R$ is written as $r_n=r_0+n$, $n=0,1,2$, where $r_0=\frac{1}{2}(1+\sqrt{g+\frac{1}{4}})$.  Whether the shift of 1 is related to this case need to be examined further.
\section{Conclusion and outlook}
We showed that the weight function of paths defined by the fixed point action can be optimized by the ESN using the $\tanh$ function for expressing the nonlinearity.

For getting the optimal solution of these problems, Machine Learning techniques can be applied. Nonlinearity and hysteresis could be explored in these basis. As shown in \cite{DSP10}, the technique is applicable for an  extension of dental investigation, as an example, which is restricted at present in (2+1)D system. The damaged position can be detected in the 3D space if receivers and transducers of TR waves are distributed in the 3D space and signals traveling to all directions can be detected. 

We showed that the (2+1)D TR-NEWS data can be simulated by conformal transformation model.


\appendix

\section{Recurrent ESN procedure}
The ESN recurrent calculation was done as follows \cite{SFDS24}.
\begin{itemize}
\item Using the ${\bf h}[4]=f(W_{ir}{\bf x}[4])$, where ${\bf x}[4]$ is produced randomly, we calculate ${\bf y}[4]=g(W_{io}{\bf x}[4]+W_{ro}{\bf h}[4])=({y}_1[4],\cdots, {y}_6[4])$, where $g$ and $f$ are the logistic sigmoid function or $\tanh$ function. 

From the second cycle, calculate ${\bf h}[4]=f(W_{rr}^{(4)}{\bf h}[3]+W_{ir}{\bf x}[4]+W_{or}{\bf y}[3])$ and ${\bf y}[4]=g(W_{io}{\bf x}[4]+W_{ro}{\bf h}[4]).$ Modify $W_{rr}^{(4)}\to W_{rr}^{(4)}-\eta D W_{rr}^{(4)}$.

\item Calculate ${\bf h}[5]=f(W_{rr}^{(5)}{\bf h}[4]+W_{i r}{\bf x}[5]+W_{or}{\bf y}[4])=({h}_1[5], {h}_2[5],\cdots,{h}_7[5])$ and ${\bf y}[5]=g(W_{io}{\bf x}[5]+W_{ro}{\bf h}[5])$. Modify $W_{rr}^{(5)}\to W_{rr}^{(5)}-\eta D W_{rr}^{(5)}$.

\item Calculate ${\bf h}[6]=f(W_{rr}^{(6)}{\bf h}[5]+W_{i r}{\bf x}[6]+W_{or}{\bf y}[5])$ and ${\bf y}[6]=g(W_{io}{\bf x}[6]+W_{ro}{\bf h}[6])$. 

Modify $W_{rr}^{(6)}\to W_{rr}^{(6)}-\eta D W_{rr}^{(6)}$.
\item Calculate ${\bf h}[7]=f(W_{rr}^{(7)}{\bf h}[6]+W_{i r}{\bf x}[7]+W_{or}{\bf y}[6])$ and ${\bf y}[7]=g(W_{io}{\bf x}[7]+W_{ro}{\bf h}[7])$.

Modify $W_{rr}^{(7)}\to W_{rr}^{(7)}-\eta D W_{rr}^{(7)}$.

\item Calculate ${\bf h}[8]=f(W_{rr}^{(8)}{\bf h}[7]+W_{i r}{\bf x}[8]+W_{or}{\bf y}[7])$ and ${\bf y}[8]=g(W_{ro}{\bf h}[8])$.

Modify $W_{rr}^{(8)}\to W_{rr}^{(8)}-\eta D W_{rr}^{(8)}$.

\item Calculate ${\bf h}[9]=f(W_{rr}^{(9)}{\bf h}[8]+W_{i r}{\bf x}[9]+W_{or}{\bf y}[8])$ and ${\bf y}[9]=g(W_{io}{\bf x}[9]+W_{ro}{\bf h}[9])$.

Modify $W_{rr}^{(9)}\to W_{rr}^{(9)}-\eta D W_{rr}^{(9)}$.
\item Calculate ${\bf h}[10]=f(W_{rr}^{(10)}{\bf h}[9]+W_{i r}{\bf x}[10]+W_{or}{\bf y}[9])$ and ${\bf y}[10]=g(W_{io}{\bf x}[10]+W_{ro}{\bf h}[10])$. Modify $W_{rr}^{(10)}\to W_{rr}^{(10)}-\eta D W_{rr}^{(10)}$.

\item Calculate ${\bf h}[11]=f(W_{rr}^{(11)}{\bf h}[10]+W_{i r}{\bf x}[11]+W_{or}{\bf y}[10])$ and ${\bf y}[11]=g(W_{io}{\bf x}[11]+W_{ro}{\bf h}[11])$.
Modify $W_{rr}^{(11)}\to W_{rr}^{(11)}-\eta D W_{rr}^{(11)}$.
\item Calculate ${\bf h}[12]=f(W_{rr}^{(12)}{\bf h}[11]+W_{i r}{\bf x}[12]+W_{or}{\bf y}[11])$ and ${\bf y}[12]=g(W_{io}{\bf x}[12]+W_{ro}{\bf h}[12])$. 

Modify $W_{rr}^{(12)}\to W_{rr}^{(12)}-\eta D W_{rr}^{(12)}$.
 
\item Calculate ${\bf h}[13]=f(W_{rr}^{(13)}{\bf h}[12]+W_{i r}{\bf x}[13]+W_{or}{\bf y}[12])$ and ${\bf y}[13]=g(W_{io}{\bf x}[13]+W_{ro}{\bf h}[13])$.
Modify $W_{rr}^{(13)}\to W_{rr}^{(13)}-\eta D W_{rr}^{(13)}$.
\item Calculate ${\bf h}[14]=f(W_{rr}{\bf h}[13]+W_{i r}{\bf x}[14]+W_{or}{\bf y}[13])$ and ${\bf y}[14]=g(W_{io}{\bf x}[14]+W_{ro}{\bf h}[14])$.
Modify $W_{rr}^{(14)}\to W_{rr}^{(14)}-\eta D W_{rr}^{(14)}$.
\item Calculate ${\bf h}[15]=f(W_{rr}^{(15)}{\bf h}[14]+W_{i r}{\bf x}[15]+W_{or}{\bf y}[14])$ and ${\bf y}[15]=g(W_{io}{\bf x}[15]+W_{ro}{\bf h}[15])$.
Modify $W_{rr}^{(15)}\to W_{rr}^{(15)}-\eta D W_{rr}^{(15)}$.

\item Calculate ${\bf h}[16]=f(W_{rr}^{(16)}{\bf h}[15]+W_{i r}{\bf x}[16]+W_{or}{\bf y}[15])$ and ${\bf y}[16]=g(W_{ro}{\bf h}[16])$.

Modify $W_{rr}^{(16)}\to W_{rr}^{(16)}-\eta D W_{rr}^{(16)}$ 

\item Calculate ${\bf h}[1]=f(W_{rr}^{(1)}{\bf h}[16]+W_{i r}{\bf x}[1]+W_{or}{\bf y}[16])$ and ${\bf y}[1]=g(W_{io}{\bf x}[1]+W_{ro}{\bf h}[1])$.

Modify $W_{rr}^{(1)}\to W_{rr}^{(1)}-\eta D W_{rr}^{(12)}$

\item Calculate ${\bf h}[2]=f(W_{rr}^{(2)}{\bf h}[1]+W_{i r}{\bf x}[2]+W_{or}{\bf y}[1])$ and ${\bf y}[2]=g(W_{io}{\bf x}[2]+W_{ro}{\bf h}[2])$.

Modify $W_{rr}^{(2)}\to W_{rr}^{(2)}-\eta D W_{rr}^{(2)}$.

\item Calculate ${\bf h}[3]=f(W_{rr}^{(3)}{\bf h}[2]+W_{i r}{\bf x}[3]+W_{or}{\bf y}[2])$ and ${\bf y}[3]=g(W_{io}{\bf x}[3]+W_{ro}{\bf h}[3])$.
 
Modify $W_{rr}^{(3)}\to W_{rr}^{(3)}-\eta D W_{rr}^{(3)}$.
\end{itemize}
{\bf Acknowledgments}:
SF thanks the Japan Industrial Science Laboratory (Nissanken) for the financial aid of the travel expense to INSA Centre Val de Loire, Blois Campus in November 2023, and Prof. M. Arai and Prof. K. Hamada for allowing the use of workstations in their laboratory.

\end{document}